# Interplay between superconductivity and magnetism in one-unit-cell LaAlO$_3$ capped with SrTiO$_3$

Yongsu Kwak,[†,‡] Woojoo Han,[†,§] Thach D. N. Ngo,[†] Dorj Odkhuu,[⊥] Jihwan Kim,[†] Young Heon Kim,[†] Noejung Park,[¶] Sonny H. Rhim,[#] Myung-Hwa Jung,[∥] Junho Suh,[†] Seung-Bo Shim,[†] Mahn-Soo Choi,[□] Yong-Joo Doh,[◊] Joon Sung Lee,[♣] Jonghyun Song,*[,‡] and Jinhee Kim*[,†]

[†]Korea Research Institute of Standards and Science, Daejeon 34113, [‡]Department of Physics, Chungnam National University, Daejeon 34134, [§]Department of Nanoscience, University of Science and Technology, Daejeon 34113, [⊥]Department of Physics, Incheon National University, Incheon 22012, [¶]School of Natural Science, Ulsan National Institute of Science and Technology, Ulsan 44919, [#]Department of Physics and Energy Harvest Storage Research Center (EHSRC), University of Ulsan, Ulsan 44610, [∥]Department of Physics, Sogang University, Seoul 04107, [□]Department of Physics, Korea University, Seoul 02841, [◊]Department of Physics and Photon Science, Gwangju Institute of Science and Technology, Gwangju 61005, [♣]Department of Display and Semiconductor Physics, Korea University Sejong Campus, Sejong 30019, Korea

ABSTRACT

To form a conducting layer at the interface between the oxide insulators $LaAlO_3$ and $SrTiO_3$, the $LaAlO_3$ layer on the $SrTiO_3$ substrate must be at least four unit-cells-thick. The $LaAlO_3/SrTiO_3$ heterointerface thus formed exhibits various intriguing phenomena such as ferromagnetism and superconductivity. It has been widely studied for being a low-dimensional ferromagnetic oxide superconducting system with a strong gate-tunable spin–orbit interaction. However, its lack of stability and environmental susceptiveness have been an obstacle to its further experimental investigations and applications. Here, we demonstrate that capping the bilayer with $SrTiO_3$ relieves this thickness limit, while enhancing the stability and controllability of the interface. In addition, the $SrTiO_3$-capped $LaAlO_3$ exhibits unconventional superconductivity; the critical current dramatically increases under a parallel magnetic field, and shows a reversed hysteresis contrary to the conventional hysteresis of magnetoresistance. Its superconducting energy gap of $\Delta \sim 1.31 k_B T_c$ also deviates from conventional BCS-type superconductivity. The oxide trilayer could be a robust platform for studying the extraordinary interplay of superconductivity and ferromagnetism at the interface electron system between $LaAlO_3$ and $SrTiO_3$.

Since its discovery by Ohtomo and Hwang[1] in 2004, the two-dimensional electron gas (2DEG) formed at the interface between the band insulators $LaAlO_3$ (LAO) and $SrTiO_3$ (STO) has attracted great interest. Scenarios based on oxygen vacancies,[2,3] cationic inter-diffusion,[4] or polar catastrophe[5] have been proposed to explain the interface conduction. According to the last scenario, it is the electrostatic potential buildup in the polar LAO layers which induces an electron transfer from the LAO surface to the embedded interface as the LAO thickness becomes equal to or thicker than 4 unit cells (uc),[6] leading to the formation of the 2DEG. The LAO/STO heterostructure provides a unique platform to investigate the coexistence and interplay of superconductivity and ferromagnetism[7–9], yet the interplay between the two strongly correlated phenomena remains challenging to understand.

Although the electron channel of LAO/STO is embedded under the LAO layer, it can be easily influenced by chemical[10] and electrostatic environment. While this environmental susceptiveness might be beneficial for gated device and sensor applications, it will be detrimental for consistent experimental researches and for device robustness in general. In this work, we study the electronic transport properties of LAO/STO capped with STO. In this trilayer system, interface conduction occurs even when the LAO layer is thinner than the critical thickness, 4 uc, as long as the total thickness of the LAO and STO films equals or larger than 4 uc. This system exhibits unconventional superconductivity and ferromagnetic ordering at low temperatures. The interplay between superconductivity and ferromagnetism is shown by a sharp increase of the critical current at non-zero magnetic field. The superconducting gap is determined by forming a tunnel-type superconducting junction with Al. The successful production of a clean capacitive vertical Josephson junction proves the firmness and stability of the STO capping layer, which will help the electron system to withstand additional fabrication processes. We also note that our sample did not show a change of properties even after several thermal cycles between the room temperature and the superconducting temperatures, exhibiting the robustness of the STO-capped LAO/STO.

The oxide heterostructures were prepared by pulsed laser deposition[11,12] (see Experimental Section). The LAO and STO layers of various thicknesses were grown on $TiO_2$-terminated STO substrates (Figure S1, Supporting Information). Figure 1a shows high-angle annular dark-field and annular bright-field scanning transmission electron microscopy images of a (STO)$_{30}$/(LAO)$_1$/STO sample. The 1-uc-thick LAO film appears clearly, with the La atoms standing out from the array of Sr atoms. Note that the LAO film is fully strained and that the interfaces are atomically sharp and free of defects such as pinholes and misfit dislocations.

The electrical transport properties were measured by the van der Pauw method. Figure 1b shows the temperature-dependent resistance curves of the (STO)$_n$/(LAO)$_1$/STO trilayers with varying $n$. Surprisingly, the 1 uc-LAO layer—though it is far thinner than the well-known critical thickness of 4 uc—induces electrical conduction when the capping STO layer is 3 uc or above. Once the trilayer becomes conductive, electrical transport properties such as sheet resistance, carrier density, and mobility become rather insensitive to the thickness of the STO layer (Figure S2). To obtain a comprehensive picture, we assessed electrical transport properties of the (STO)$_n$/(LAO)$_m$/STO trilayer as functions of $m$ and $n$. Figure 1c shows the room-temperature conductivity of the trilayer system. The trilayer becomes conductive when $m + n \geq 4$.

Our experimental observations were confirmed by theoretical calculations (see Experimental Section). For the LAO/STO bilayers, the energy bandgap decreases with increasing LAO thickness and vanishes as it reaches 4 uc (Figure 1d). The electron channel forms at the interface and the hole channel at the surface, supporting the well-known polar catastrophe scenario.[5,13,14] The 1 uc-LAO on STO substrate is insulating, but adding the STO capping layer dramatically changes the electronic band structure of the system. The bandgap decreases as the STO thickness increases and eventually vanishes at 3 uc of STO, which enables electron transfer from the surface to the interface (Figure S3). This charge transfer causes the insulator-to-metal transition of the oxide system. Our result partly coincides with those from previous experimental and theoretical studies on the STO capping and the LAO/STO superlattice. However, our work further broadens the horizon by redefining the condition for conduction channel formation, and by revealing its superconducting properties related to ferromagnetism as well.

A 1 uc-LAO capped with STO exhibits superconductivity at low temperatures (Figure S4). For the electrical measurements, we fabricated a Hall bar-type mesa with a $(STO)_7/(LAO)_1/STO$ trilayer by using photolithography and ion milling (see Figure 2a and Experimental Section). At temperatures below $T_c$ = 180 mK, the interface resistance drops to zero, and its current–voltage ($I$–$V$) curves show a supercurrent branch. At $T$ = 20 mK, the critical current is $I_c$ = 1.2 µA (Figure S4a) and the critical magnetic field is $H_{c\perp}$ = 1.4 kOe ($H_{c\|}$ = 5.0 kOe) for the field perpendicular (parallel) to the interface. Our $H_{c\perp}$ is similar to the value observed in a LAO/STO bilayer system,[15] while our $H_{c\|}$ is three times smaller. By assuming a paramagnetic pair-breaking effect in a two-dimensional superconductor, an effect known as the Pauli limit,[16] we can estimate the superconducting gap of the trilayer system to be $\Delta = g\mu_B H_{c\|}/\sqrt{2} = 41\,\mu eV$, where $g$ is the electron $g$-factor and $\mu_B$ is the Bohr magneton.

Figure 2b shows how the $I$–$V$ characteristics change with the parallel magnetic field. Surprisingly, $I_c$ sharply increases at non-zero magnetic field. Moreover, the field-induced $I_c$ modulation shows clear advancing hysteresis, as depicted in Figure 2c. When the field is swept from positive to negative (or vice versa), the maximum $I_c$ occurs at +75 (or −75) Oe. The field-induced $I_c$ enhancement and its hysteretic behavior suggest an interplay between the superconductivity and ferromagnetism.

Another hysteretic behavior appears in the magnetoresistance (MR) curve, as shown in Figure 2d. This MR hysteresis, an evidence of ferromagnetic ordering, turns out to be quite sensitive to the field sweep rate. The MR peak becomes sharper and narrower at slower sweep rates and eventually vanishes (Figure S4b). A similar feature has appeared in a LAO/STO bilayer system[17] and was suggested as evidence of an interplay between superconductivity and ferromagnetism. Our observations, however, reveal that the interface superconductivity is not a necessary condition for the MR hysteresis, although the superconducting fluctuation near $T_c$ can enhance its amplitude; in our trilayer system, the MR hysteresis persists up to $T \approx 800$ mK, far above $T_c$ (Figure S4c). Note that the MR hysteresis resembles the conventional retarding hysteresis of $M$–$H$, while the aforementioned $I_c$ hysteresis occurs in the reverse order. The physical origin of this anomalous $I_c$ hysteresis will be discussed later.

To gain more quantitative information on the interface superconductivity of the STO–capped (LAO)$_1$/STO system, a tunnel-type Josephson junction was fabricated by depositing a Ti (10 nm)/Al (140 nm) superconducting electrode on the oxide Hall bar (see Figure 2a and Experimental Section). Figure 3a shows the $I$–$V$ curves of the tunnel junction at various temperatures. The junction critical current and normal resistance at 20 mK are $I_c^t = 85$ nA and $R_n = 140$ Ω, respectively. The $I$–$V$ curves show hysteresis at low temperatures, a typical behavior of a tunnel-type Josephson junction with large capacitive coupling.

In an asymmetrical Josephson junction, of which the two superconductors have different superconducting gaps, $\Delta_1$ and $\Delta_2$ ($\Delta_1 < \Delta_2$), the jumping voltage at $I_c^t$ corresponds to $\Delta_1/e$,[18] where $e$ is the elementary charge. From the $I$–$V$ curve we obtained $\Delta_1 = 20.3$ μeV at $T = 20$ mK, which is about half of the value estimated earlier from $H_{c\parallel}$. The gap ratio is given by $\Delta_1/k_B T_c = 1.31$, which deviates from the theoretical value for a BCS-type superconductor, 1.75,[19] and also from the experimental value measured from an LAO/STO bilayer.[20]

Figure 3b shows the temperature dependences of the critical current ($I_c^t$) and retrapping current ($I_r^t$), revealing that the hysteresis in the junction I–V curves vanishes near $T = 90$ mK. At higher temperatures, the differential conductance (dI/dV) curve of a Josephson junction may exhibit a series of peaks attributed to the Andreev reflections (AR). In a highly asymmetric junction ($\Delta_1 \ll \Delta_2$) as in our case, the AR peaks can appear at bias voltages of $\Delta_1/e$, $\Delta_1/2e$, $\Delta_1/3e$, ….[21] Figure 3c shows a prominent dI/dV peak corresponding to a single AR at $V = \Delta_1/e = 12$ μV at $T = 140$ mK. The temperature dependence of $\Delta_1$ can be obtained from the peak voltages of the dI/dV–V curves (Figure S5). Figure 3d shows $\Delta_1$ estimated from the voltage jump in the I–V curve (open circle) and the peak voltages of the dI/dV–V curves (filled circle). Fitting the data to the empirical formula $\Delta_1(T) = \Delta_1(0) \tanh[\alpha(T_c/T - 1)^{1/2}]$ gives $\alpha = 1.12$ (solid line), which deviates from $\alpha = 1.74$ (dashed) expected from BCS theory.[19] The measured $I_c^t$ data in Figure 3b also fit well to the Ambegaokar–Baratoff relation,[18] in which non-BCS-type $\Delta_1(T)$ for the superconducting phase at the interface and BCS-type $\Delta_2(T)$ for Al are supposed.

Both the reversed order of the $I_c$ hysteresis and the non-BCS-type superconductivity may imply an unconventional superconductivity in the STO/LAO/STO trilayer system. As one of many possibilities, we conjecture that the observed unexpected superconducting properties come from a so-called helical superconducting phase.[22,23] It is known that superconductivity in a two-dimensional electron system with strong Rashba spin−orbit coupling (SOC) can be enhanced by applying an in-plane magnetic field, which generates a spatially modulated superconducting order parameter. Such a helical superconducting phase may explain the reversed $I_c$ hysteresis observed in the trilayer system. Suppose the applied magnetic field is being decreased from a value large enough to suppress the interface superconductivity (Figure S6). As the field is decreased below $H_{c\parallel}$, the interface becomes superconducting. The superconductivity can be enhanced under around a specific magnetic field, at which the helical superconductivity is most favored. It is quite probable that the *dI/dV* peak around ~75 Oe in Figure 2b is owing to that enhancement of superconductivity. Decreasing the field further past zero should diminish and then enhance the superconductivity again. However, changing to a negative field causes the magnetization of the interface to turn around and realign with the applied field. Because the applied field is not strong enough, the realignment will create many magnetic domains which can cause superconductivity fluctuations and prevent superconductivity enhancement. Considering the strong Rashba SOC in the LAO/STO system,[24] the helical superconductivity can be a strong candidate to explain all our experimental observations. We believe that more extensive studies, both experimental and theoretical, are required to elucidate the physics behind the observations.

In this work, we have demonstrated that an electrical conduction channel is introduced in the STO/LAO/STO system when the total thickness of the LAO film and STO capping layer equals or exceeds 4 uc, and that the channel formation is associated with a charge transfer from the STO surface to the interface. This trilayer system is quite stable against thermal cycling and moisture, making itself suitable for various device applications. The electrical transport measurements reveal that the electron channel exhibits unconventional non-BCS type superconductivity. The unusual $I_c$ hysteresis observed in this system can be considered as a stark manifestation of the interplay between ferromagnetism and superconductivity. Along with the non-BCS-type energy gap, the reversed $I_c$ hysteresis requires a nontrivial theoretical model capable of providing an appropriate interpretation. If helical superconductivity is indeed responsible for the above observations, it may be possible that the trilayer gives rise to a *p*-wave topologically superconducting state, which is useful for pursuing Majorana fermions in condensed matter systems.[25,26]

**Experimental Section.** *Oxide heterostructure preparation and Hall measurements.* $TiO_2$-terminated $SrTiO_3$ (STO) substrate was prepared by selectively etching SrO by using buffered oxide etch (BOE). Before film growth, the STO substrates were annealed at 950 °C under an oxygen pressure of $2 \times 10^{-5}$ Torr for 2 h. $LaAlO_3$ (LAO) film of 1–4 unit cells (uc) and then STO capping layers of 1–10 uc were deposited on the STO substrate at 750 °C under an oxygen pressure of $10^{-5}$ Torr. The laser pulse energy was 120 mJ with a repetition rate of 1 Hz. The film growth was monitored by using reflection high-energy electron diffraction (RHEED) (Figure S1). After growth, the samples were annealed *in situ* in an oxygen-rich atmosphere (500 mTorr) at 750 °C for 30 min and then cooled to room temperature under the same oxygen pressure. The electrical properties of the unpatterned oxide heterostructures were measured using the van der Pauw method (Quantum Design PPMS). For the Hall measurements (Figure S2), an external magnetic field perpendicular to the oxide interface plane was swept from –1 T to 1 T.

*Junction fabrication and superconductivity measurements.* A Hall bar-shaped mesa with dimensions of 5 μm × 20 μm was lithographically defined on a $(STO)_7/(LAO)_1/STO$ trilayer sample. To prevent any current leakage, the exposed edges of the conducting interface in this structure were covered by hardened PMMA (polymethyl methacrylate). An Al (140 nm)/Ti (10 nm) top electrode was then deposited on the oxide Hall bar by rf sputtering to make a vertical superconducting tunnel junction (Figure 2a). To measure the *I*–*V* curves for the interface channel and the tunnel junction, we used the conventional four-point dc technique. To acquire the *dI*/*dV*–*V* curves, we used an ac differential technique.

*Band calculations.* Density functional theory (DFT) simulations were performed using the projector augmented-wave formalism, as implemented in the Vienna *ab initio* simulation package (VASP) with the generalized gradient approximation (GGA) for the exchange correlation potential. We modeled a slab consisting of 1–4 uc LAO layers ($m$) and 1–3 uc STO capping layers ($n$) stacked on a 4 uc STO substrate, where the cases for $(STO)_0/(LAO)_4/STO$ and $(STO)_3/(LAO)_1/STO$ are shown in Figure S3a. The experimental lattice constant (0.3905 nm) of STO was used for the in-plane lattice. Dipole corrections were included to eliminate any artificial electric field caused by supercell configurations. An energy cutoff of 600 eV and a 15 × 15 × 1 $k$-point grid are used in the calculations, where a force criteria of $10^{-2}$ eV/Å is employed for atomic relaxations.

The calculated energy band gaps ($E_g$) of $(STO)_n/(LAO)_m/STO$ as functions of $n$ and $m$ are shown in Figure S3b. For $(LAO)_m/STO$, $E_g$ decreases as $m$ increases, and it eventually vanishes for $m = 4$, which is the critical thickness of the insulator-to-metal transition observed in experiments so far. It is verified that a similar phenomenon occurs for $(STO)_n/(LAO)_m/STO$ when $m + n = 4$. Capping with STO dramatically affects the calculated band gaps. Notably, even 1 uc of LAO will induce metallicity if the STO capping layer is 3 uc or thicker. The layer-resolved density of states (DOS) with and without STO capping is shown in Figure S3c. Without STO capping, owing to an internal electric field, the O $2p$ state in the LAO film gradually shifts towards the Fermi level with going closer to the interface, which is consistent with previous studies.[27] For the STO-capped LAO/STO, charge transfer occurs from the surface $TiO_2$ to the interface Ti $d_{xy}$. This supports the experimental observation: a Ti atom at the interface exhibits fractional occupation of $t_{2g}$ orbital states of the $Ti^{3+}$ ($3d^1$) state. The presence of built–in potential within the thinner LAO films shifts the valence band of the STO capping layers, particularly for the $TiO_2$ layers, toward the Fermi level (Figure S3c). This behavior occurs because there are two distinct types of interfaces: an $n$-type $LaO/TiO_2$ interface and a $p$-type $SrO/AlO_2$ interface.

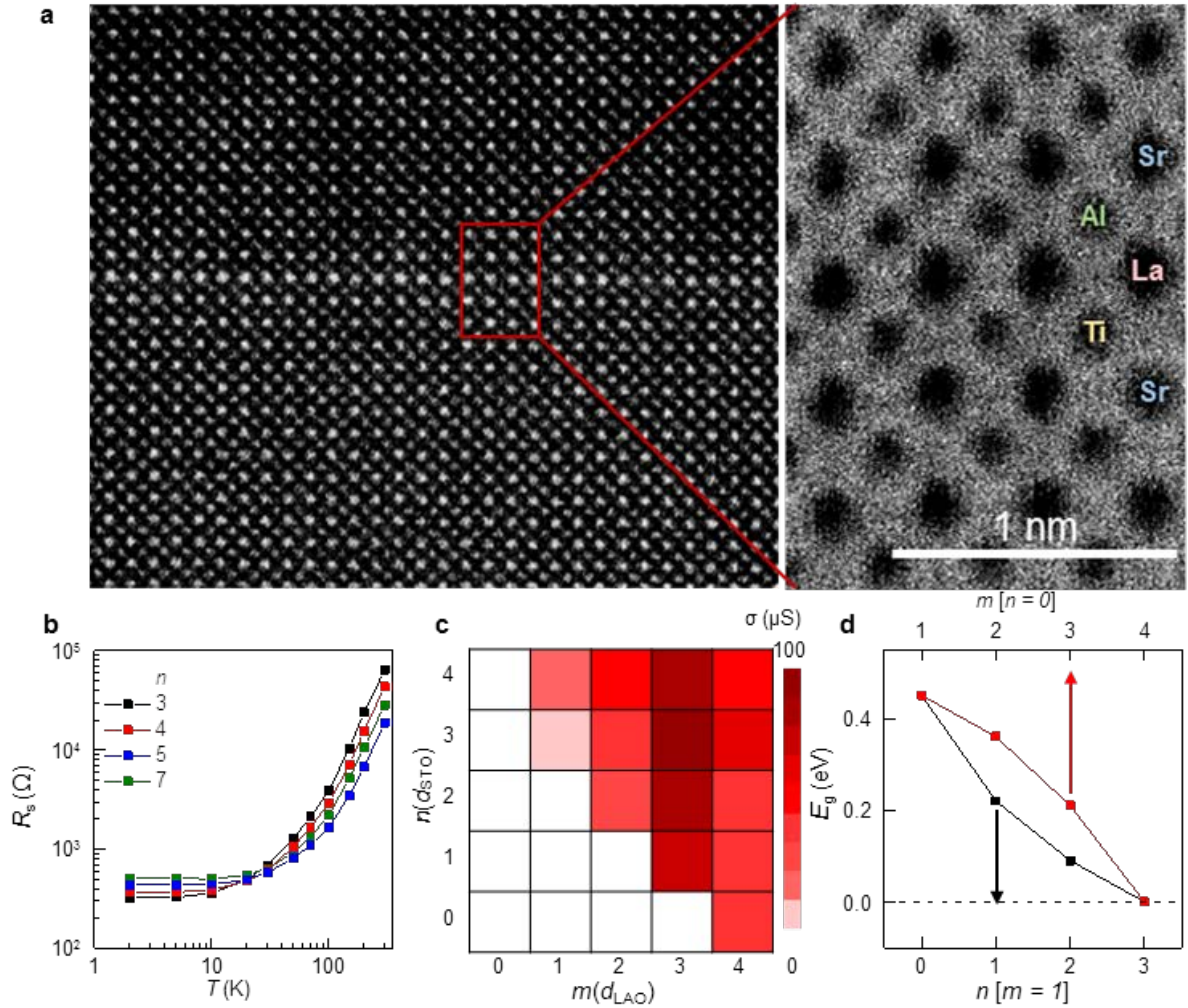

**Figure 1.** Effect of SrTiO$_3$ cap on the LaAlO$_3$/SrTiO$_3$ bilayer. (a) Cross-sectional HR-TEM image of an (SrTiO$_3$)$_{30}$/(LaAlO$_3$)$_1$/SrTiO$_3$ trilayer. A 1-uc-thick LaAlO$_3$ layer stands out from the SrTiO$_3$ background of the substrate and the capping layer. (b) The temperature dependence of sheet resistance $R_s$ for (SrTiO$_3$)$_n$/(LaAlO$_3$)$_1$/SrTiO$_3$ trilayers with various thicknesses of the capping layer $n$. (c) Interface conductance at room temperature as a function of the thicknesses of the LaAlO$_3$ ($m$) and SrTiO$_3$ ($n$) layers. (d) Calculated bandgap energy $E_g$, from density functional theory, as a function of $m$ and $n$.

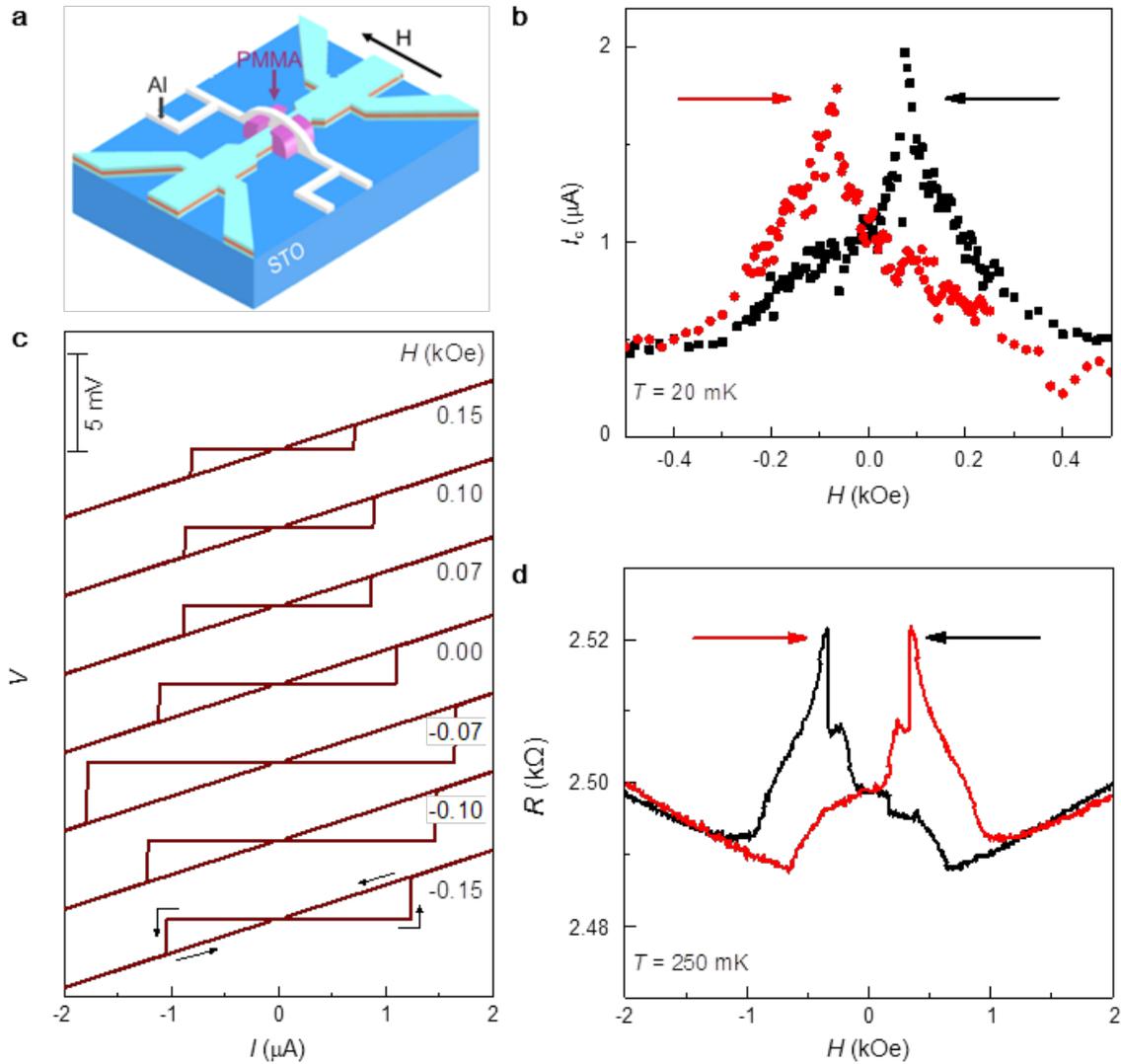

**Figure 2.** Unconventional superconductivity in the oxide heterostructure. (a) Schematic of the Al/Ti/(SrTiO$_3$)$_7$/(LaAlO$_3$)$_1$/SrTiO$_3$ tunnel junction. (b) I–V characteristics of the patterned interface with various parallel magnetic fields at $T = 20$ mK. (c) The superconducting critical current, $I_c$, as a function of magnetic field. The arrows indicate the sweep direction of the magnetic field. (d) The magnetoresistance curve of the interface measured at $T = 250$ mK with a bias current of 100 nA.

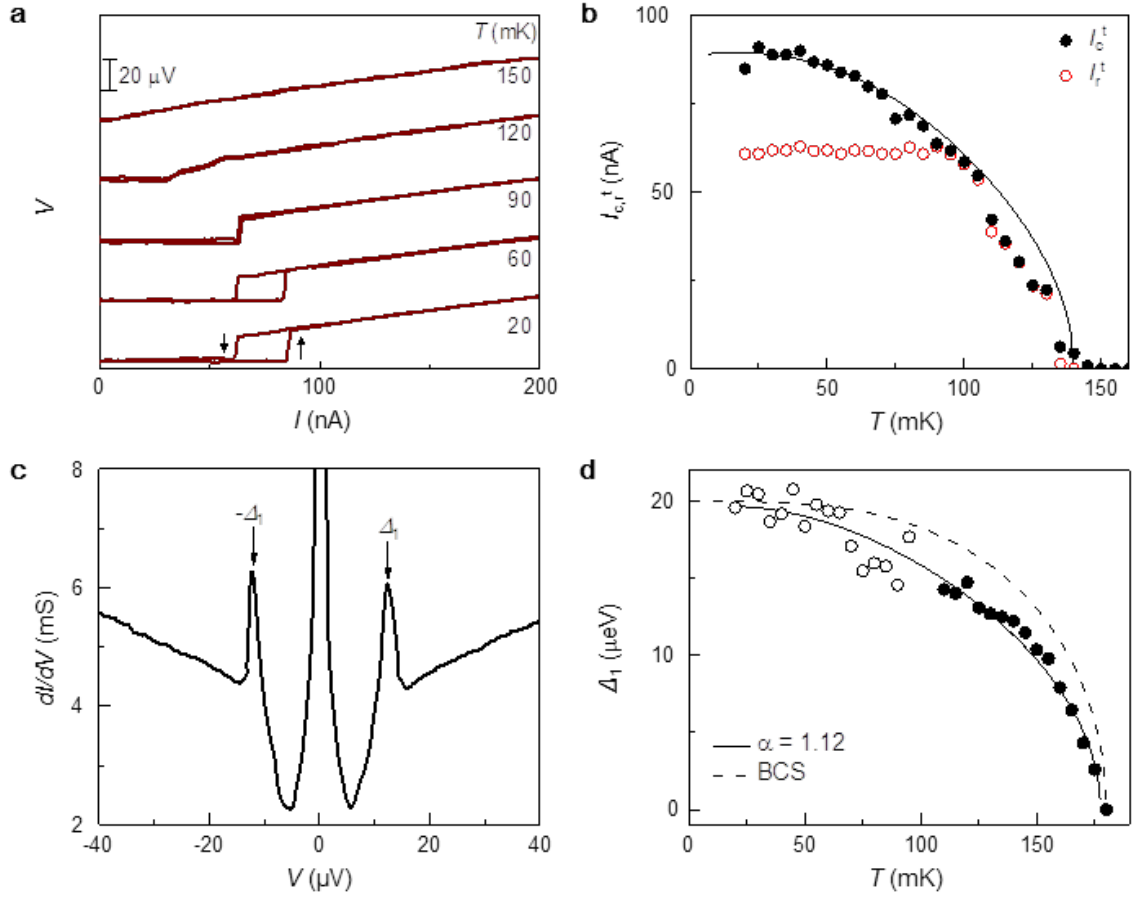

**Figure 3.** Tunnel spectroscopy of the trilayer superconducting tunnel junction. (a) Temperature-dependent $I$–$V$ characteristics of the Al/Ti/(SrTiO$_3$)$_7$/(LaAlO$_3$)$_1$/SrTiO$_3$ tunnel junction. (b) Temperature dependence of the critical current ($I_c^t$, filled circle) and the retrapping current ($I_r^t$, open circle) of the tunnel junction. The solid line is the fit to the Ambegaokar–Baratoff relation. (c) The $dI/dV$–$V$ curve of the tunnel junction at 140 mK, showing a clear peak at voltages of ±12 μV. (d) Temperature dependence of the superconducting energy gap of the interface superconductor $\Delta_1$, estimated from the voltage jump in the $I$–$V$ curve (open circle) and the peak position in the $dI/dV$–$V$ curve (filled circle). The solid line is the best fit to an empirical formula with $\alpha = 1.12$ and $\Delta_1(0)/k_B T_c = 1.31$, while the dashed line is the fit to the BCS formula with $\alpha = 1.74$.

ASSOCIATED CONTENT

**Supporting Information**.

RHEED intensity oscillations during growth of the oxide heterostructure, electrical properties of the capped LaAlO$_3$/SrTiO$_3$ interface, density functional theory study of (LaAlO$_3$)$_m$/SrTiO$_3$ and (SrTiO$_3$)$_n$/(LaAlO$_3$)$_m$/SrTiO$_3$, superconducting $I$–$V$ curves and magnetoresistance of the oxide interface, $dI/dV$–$V$ curves for the superconducting tunnel junction, a schematic model of the reversed $I_c$ hysteresis (PDF)

AUTHOR INFORMATION

**Corresponding Author**

*E-mail: songjonghyun@cnu.ac.kr

*E-mail: jinhee@kriss.re.kr

**Author Contributions**

Y. K. and W. H. contributed equally to this work. Y.K and W.H prepared the sample. W.H investigates oxide films properties. Y.K performed junction fabrication. Y.K and W.H performed the transport measurements. J.K, S.B.S, J.S help to use dilution refrigerator. Y.H.K performed transmission electron microscopy. D.O, N.P and S.H.R provide DFT calculation. M.S.C gives theoretical model for helical superconductivity. M.H.J performed magnetic property measurements. T.D.N.N, Y.J.D, J.S and J.K give analysis for experimental discussion and wrote manuscript.

**Notes**

The authors declare no competing financial interest.

ACKNOWLEDGMENT

J. Song was supported by the NRF of Korea (Grant No. 2016R1A2B4008706), Y. Doh by the SRC Center for Quantum Coherence in Condensed Matter (Grant No. 2016R1A5A1008184), and M.-S. Choi by the NRF of Korea (Grant No. 2015-003689) and by the BK21 Plus Initiative.

ABBREVIATIONS

LAO, $LaAlO_3$; STO, $SrTiO_3$; 2DEG, two-dimensional electron gas; uc, unit cells; SOC, spin−orbit coupling.

Supporting Information:

Interplay between superconductivity and magnetism in one-unit-cell LaAlO$_3$ capped with SrTiO$_3$

*Yongsu Kwak, Woojoo Han, Thach D. N. Ngo, Dorj Odkhuu, Jihwan Kim, Young Heon Kim, Noejung Park, Sonny H. Rhim, Myung-Hwa Jung, Junho Suh, Seung-Bo Shim, Mahn-Soo Choi, Yong-Joo Doh, Joon Sung Lee, Jonghyun Song,\* and Jinhee Kim\**

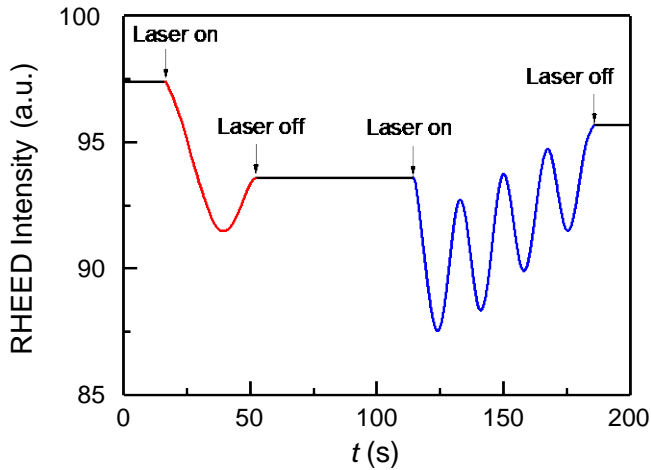

**Figure S1. Oxide heterostructure growth.** RHEED intensity oscillations during the growth of the LaAlO$_3$ film and SrTiO$_3$ capping layer on the TiO$_2$-terminated SrTiO$_3$ substrate.

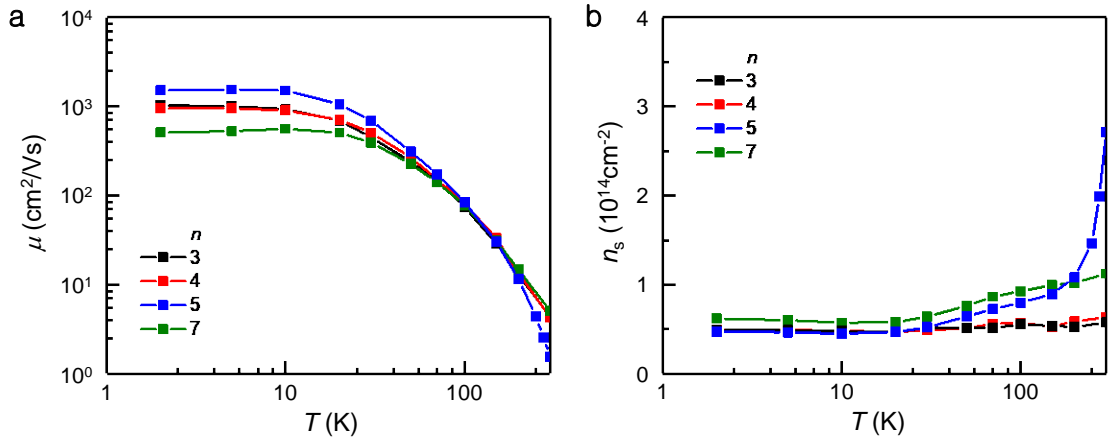

**Figure S2. Electrical properties of the capped LaAlO$_3$/SrTiO$_3$ interface.** Temperature dependence of the (a) carrier mobility $\mu$, and (b) carrier density $n_s$ in the (SrTiO$_3$)$_n$/(LaAlO$_3$)$_1$/SrTiO$_3$ trilayers with varying capping layer thickness $n$.

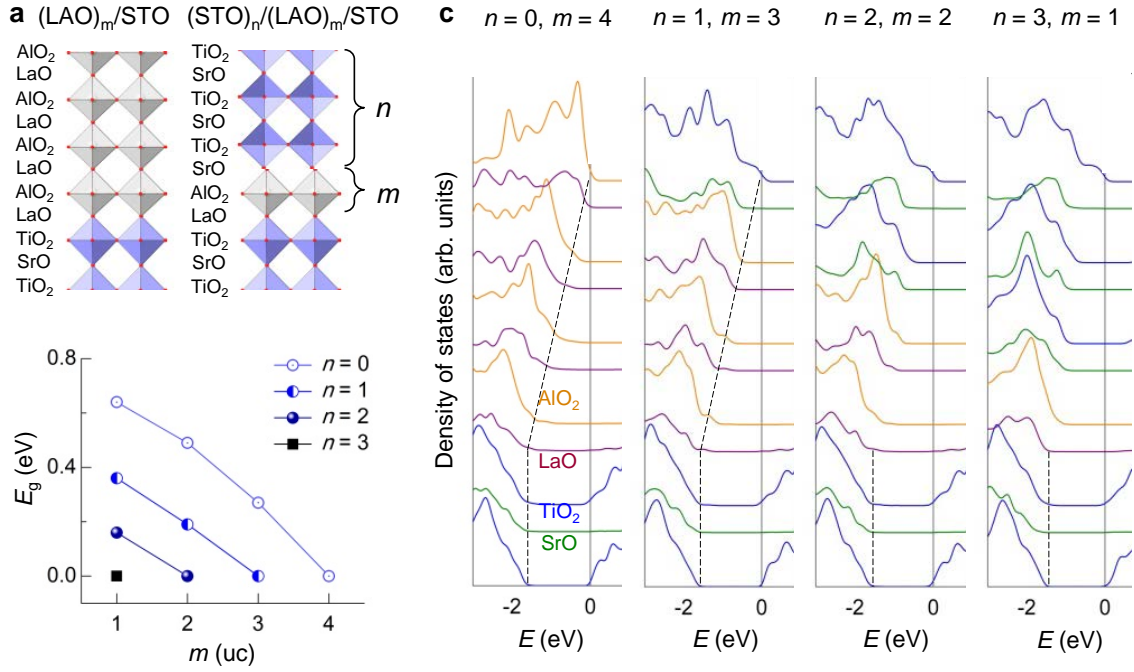

**Figure S3. Density functional theory study.** (a) Schematic atomic structures of $(LaAlO_3)_m/SrTiO_3$ (left) and $(SrTiO_3)_n/(LaAlO_3)_m/SrTiO_3$ (right), where $m$ and $n$ are the numbers of $LaAlO_3$ and $SrTiO_3$ unit cell layers, respectively. Blue and gray octahedra represent $SrTiO_3$ and $LaAlO_3$, respectively. Small red spheres at the vertices of the octahedron are oxygen atoms. For simplicity, cations and the bottom three unit cells of the $SrTiO_3$ substrate are not shown. (b) Energy bandgap $E_g$ of $(SrTiO_3)_n/(LaAlO_3)_m/SrTiO_3$ as a function of $n$ and $m$. (c) Layer-resolved density of states for $(SrTiO_3)_n/(LaAlO_3)_m/SrTiO_3$ at the critical thickness for the insulator-to-metal transition ($m + n = 4$). The Fermi level is set to zero energy.

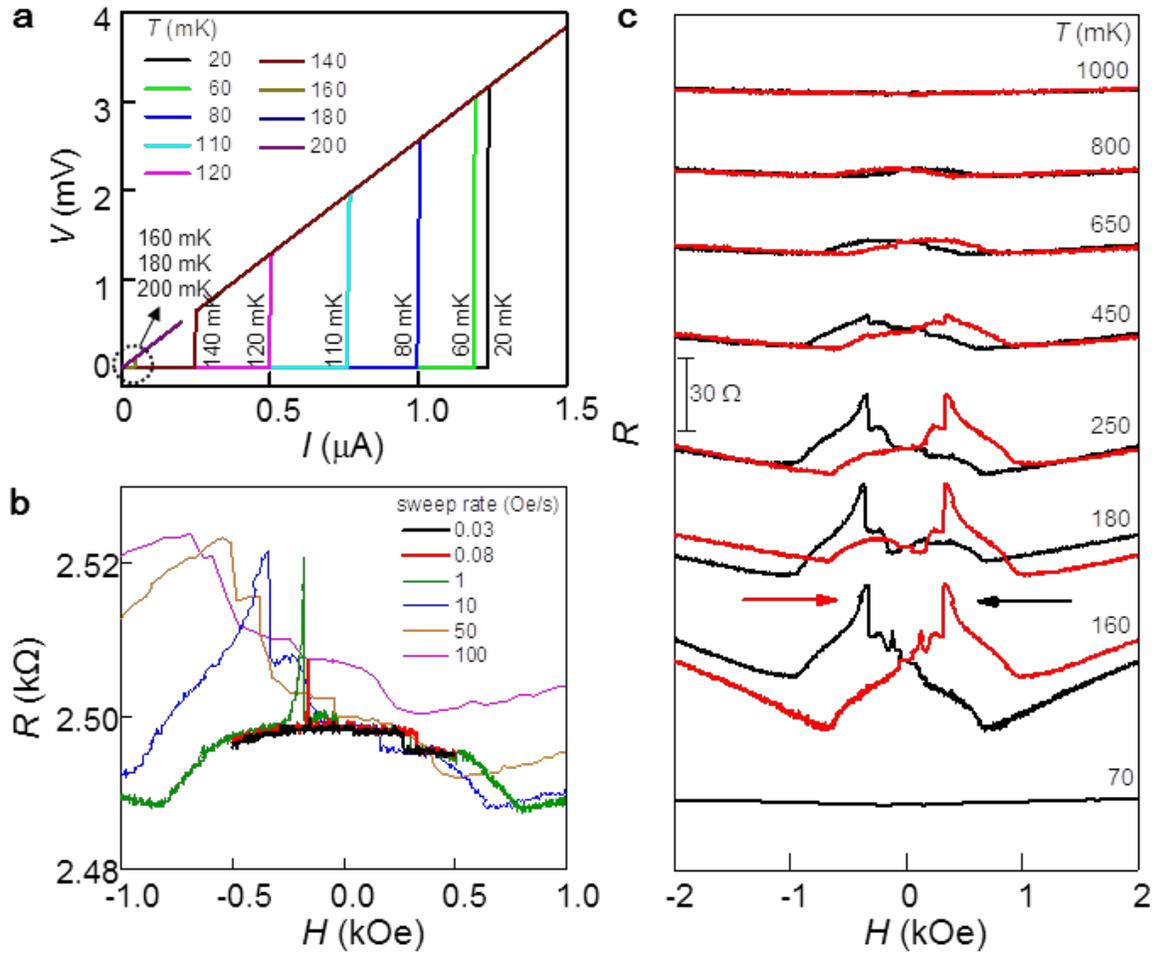

**Figure S4. Superconductivity and magnetoresistance of the oxide interface.** (a) The evolution of *I–V* characteristics with temperature. (b) The MR curve at 250 mK with varying magnetic field sweep rate. For simplicity, the sweep direction of the magnetic field is fixed. (c) The temperature dependence of the MR curve. The external magnetic field is applied parallel to the interface. The field sweep rate is fixed to 10 Oe/s.

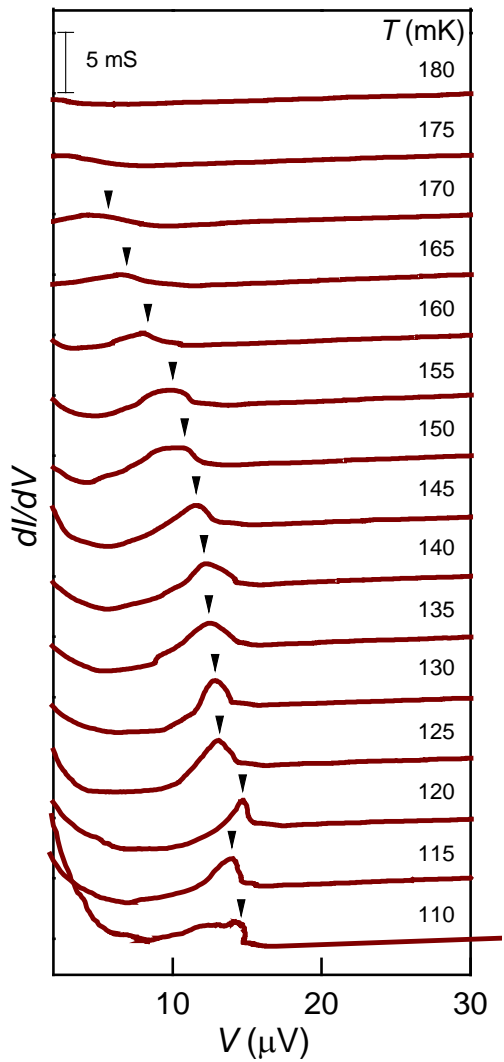

**Figure S5. Tunnel spectroscopy.** Temperature dependence of the differential conductance–voltage, *dI/dV–V*, curves for the tunnel junction. The peaks in *dI/dV* owing to the Andreev reflection are marked with arrows.

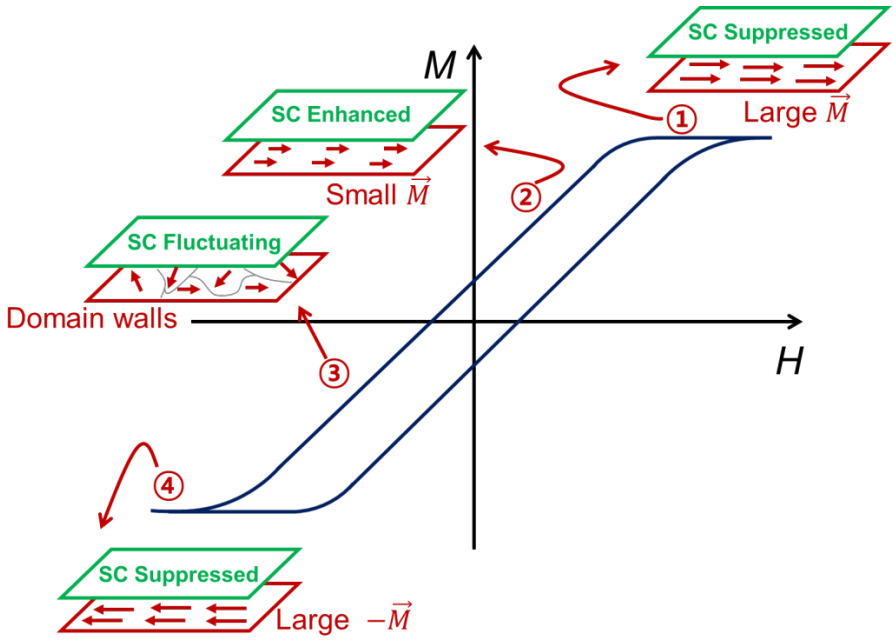

**Figure S6. A schematic model of the reversed $I_c$ hysteresis,** accounting for the possible interplay of the helical superconducting state and the ferromagnetic ordering.